\documentclass[sn-mathphys-ay]{sn-jnl}

\usepackage{graphicx}%
\usepackage{multirow}%
\usepackage{amsmath,amssymb,amsfonts}%
\usepackage{amsthm}%
\usepackage{mathrsfs}%
\usepackage[title]{appendix}%
\usepackage{xcolor}%
\usepackage{textcomp}%
\usepackage{manyfoot}%
\usepackage{booktabs}%
\usepackage{algorithm}%
\usepackage{algorithmicx}%
\usepackage{algpseudocode}%
\usepackage{listings}%

\usepackage{booktabs}
\usepackage{makecell}
\usepackage{subfigure}
\usepackage{natbib}
\usepackage[figuresright]{rotating}
\usepackage{amsmath}
\usepackage{hyperref}
\makeatletter
\def\UrlAlphabet{%
      \do\a\do\b\do\c\do\d\do\e\do\f\do\g\do\h\do\i\do\j%
      \do\k\do\l\do\m\do\n\do\o\do\p\do\q\do\r\do\s\do\t%
      \do\u\do\v\do\w\do\x\do\y\do\z\do\A\do\B\do\C\do\D%
      \do\E\do\F\do\G\do\H\do\I\do\J\do\K\do\L\do\M\do\N%
      \do\O\do\P\do\Q\do\R\do\S\do\T\do\U\do\V\do\W\do\X%
      \do\Y\do\Z}
\def\UrlDigits{\do\1\do\2\do\3\do\4\do\5\do\6\do\7\do\8\do\9\do\0}
\g@addto@macro{\UrlBreaks}{\UrlOrds}
\g@addto@macro{\UrlBreaks}{\UrlAlphabet}
\g@addto@macro{\UrlBreaks}{\UrlDigits}
\makeatother

\usepackage{enumerate}

\setcitestyle{aysep={}}

\theoremstyle{thmstyleone}%
\theoremstyle{thmstyletwo}%

\theoremstyle{thmstylethree}%

\begin{document}

\title[Article Title]{Toward Efficient Testing of Graph Neural Networks via Test Input Prioritization}


\author[1]{\fnm{Lichen} \sur{Yang}}\email{lcyang.cn@gmail.com}

\author[1]{\fnm{Qiang} \sur{Wang}}\email{qiang.wang@hit.edu.cn}
\author[3]{\fnm{Zhonghao} \sur{Yang}}\email{ashzhonghao@gmail.com}

\author*[1]{\fnm{Daojing} \sur{He}}\email{hedaojinghit@163.com}

\author*[2]{\fnm{Yu} \sur{Li}}\email{li.yu@zju.edu.cn}

\affil[1]{\orgdiv{Computer Science and Technology}, \orgname{Harbin Institute of Technology (Shenzhen)}, \orgaddress{\city{Shenzhen}, \postcode{518000}, \state{Guangdong}, \country{China}}}

\affil[2]{\orgdiv{Integrated Circuits}, \orgname{Zhejiang University}, \orgaddress{\city{Hangzhou}, \postcode{310058}, \state{Zhejiang}, \country{China}}}

\affil[3]{\orgdiv{Software Engineering Institute}, \orgname{East China Normal University}, \orgaddress{\street{North Zhongshan Road}, \postcode{200061}, \state{Shanghai}, \country{China}}}


\abstract{
Graph Neural Networks (GNNs) have demonstrated remarkable efficacy in handling graph-structured data; however, they exhibit failures after deployment, which can cause severe consequences.
Hence, conducting thorough testing before deployment becomes imperative to ensure the reliability of GNNs. 
However, thorough testing requires numerous manually annotated test data. 
To mitigate the annotation cost, strategically prioritizing and labeling high-quality unlabeled inputs for testing becomes crucial, which facilitates uncovering more model failures with a limited labeling budget.
Unfortunately, existing test input prioritization techniques either overlook the valuable information contained in graph structures or are overly reliant on attributes extracted from the target model, \textit{i.e., model-aware attributes}, whose quality can vary significantly.
To address these issues, we propose a novel test input prioritization framework, named \textit{GraphRank}, for GNNs. 
GraphRank introduces model-agnostic attributes to compensate for the limitations of the model-aware ones. 
It also leverages the graph structure information to aggregate attributes from neighboring nodes, thereby enhancing the model-aware and model-agnostic attributes. 
Furthermore, GraphRank combines the above attributes with a binary classifier, using it as a ranking model to prioritize inputs. 
This classifier undergoes iterative training, which enables it to learn from each round's feedback and improve its performance accordingly. Extensive experiments demonstrate GraphRank's superiority over existing techniques.}




\maketitle

\section{Introduction}
Graph Neural Networks (GNNs)~\citep{kipf2017semi, velivckovic2018graph, wu2020comprehensive} have proven to be effective for various graph-based applications. The tasks of GNNs are primarily divided into three levels: node level, edge level, and graph level. Among these, node classification is a common task at the node level, which can be applied to various scenarios. For example, it can predict the research field of papers in citation networks~\citep{tang2015line, sen2008collective}. In the protein-protein interaction network, each node can be assigned several gene ontology types~\citep{subramanian2005gene, xu2006discovering}. In social networks, it can be employed to predict community structure~\citep{girvan2002community, rosvall2007information}.

Similar to other Deep Neural Networks (DNNs), GNNs may also make mistakes, caused by factors such as distribution shifts between the training and testing data, limitations in model capability, and/or insufficient training data.
Therefore, testing the quality of GNNs is essential to ensure their reliability. However, given the scale and complexity of current graphs, the cost of manually labeling a large number of test inputs is prohibitively high, which hinders the testing process. 

    To address this problem, test input prioritization has been introduced~\citep{byun2019input, feng2020deepgini, dang2023graphprior}.
    For a given target model, \textit{i.e.}, the under-test model, they attempt to estimate the priority scores for each test input and select inputs with higher scores under a given labeling budget. 
    The score indicates the probability of the target model making an incorrect prediction.  
    Thus, the selected inputs are more likely to reveal more failures within the limited labeling budget, improving the test efficiency. 
    These inputs can subsequently facilitate targeted model repair by guiding engineers in identifying and addressing specific model deficiencies. Consequently, test input prioritization serves as an effective approach to mitigating model failures.
    However, the majority of existing techniques~\citep{ma2018deepgauge, byun2019input, feng2020deepgini, wang2021prioritizing} are designed for traditional Deep Neural Networks (DNNs) which handle non-graph data (e.g., images). 
    In the context of GNN testing, these methods treat nodes as isolated entities, which restricts their ability to fully exploit the information from the graph structure.

    To improve the test efficiency for GNNs, a few tailored test input prioritization techniques have been proposed~\citep{dang2023graphprior, li2024test}.
    These methods focus on node classification tasks and aim to prioritize unlabeled nodes for testing.
    These methods are based on the concept of mutation testing and generate various mutated models. For each unlabeled node, they estimate the priority score by analyzing the prediction differences between the target model and the mutants.
Although these methods can be applied to GNNs, they rely on attributes derived from the target model—specifically, model predictions, which we refer to as \textit{model-aware attributes}. The reliability of these attributes is significantly influenced by the quality of the target model. In cases where the target model is of low quality, correctly classified nodes may still receive disproportionately high priority scores. This discrepancy arises because the predictions from both the low-quality model and its mutants lack reliability, resulting in flawed prioritization when solely depending on them.

To address the above issues, this paper presents a novel test input prioritization framework, namely \textit{GraphRank}, to facilitate the testing efficiency for node classification tasks in GNNs. On the one hand, the limited reliability of model-aware attributes motivates us to explore \textit{model-agnostic attributes}, \textit{i.e.}, the attributes that are independent of the target model and only related to the graph data itself. On the other hand, the properties of graph structure inspire us to design a mechanism that can fully leverage the valuable graph structure information. Therefore, GraphRank seamlessly combines model-aware and model-agnostic attributes (e.g., node features) along with graph structural insights to achieve a high level of test efficiency. 
    Compared with the model-aware attributes, model-agnostic attributes offer supplementary information that can help reduce the potential negative impact arising from the quality of the target model.
    Concurrently, most graphs exhibit homophily—where most nodes share similar properties with their neighbors—highlighting the importance of graph structure in leveraging information from adjacent nodes.

    The implementation of GraphRank involves three steps. First, we explore both model-aware and model-agnostic attributes. Model-aware attributes are extracted from the output of a target GNN model, e.g., deterministic output and probabilistic output attributes. Model-agnostic attributes are independent of the model and focus on graph data information, e.g., graph node and node degree attributes. 
    Second, we enhance these attributes by leveraging the graph structure information to aggregate neighbor attributes. Finally, we integrate the above attributes with a learnable binary classifier, which serves as the ranking model to predict priority scores for unlabeled nodes, enabling their prioritization. 
    To optimize the performance of the classifier, we split the labeling budget into multiple portions and train the classifier iteratively. In each training iteration, a portion of the budget is used to annotate high-priority nodes. These newly labeled nodes are incorporated into the training set of the classifier, conducting the next round of training. The iterative training strategy allows the classifier to receive feedback about prioritization quality and make corresponding adjustments.

    In conclusion, our contributions are summarized as follows:
    \begin{itemize}

        \item We design a test input prioritization framework, namely GraphRank, for node classification tasks in GNNs. GraphRank explores various types of attributes and graph structure information for test input prioritization, making it effective for testing GNNs of different qualities. Additionally, our framework is flexible, allowing for the integration of further effective attributes as needed.  
       
        \item In our framework, we introduce both model-aware and model-agnostic attributes. Model-aware attributes include deterministic and probabilistic outputs from the target GNN model, while model-agnostic attributes contain graph data features such as node features and node degrees. The model-agnostic attributes can compensate for the drawback of model-aware attributes limited by the variation of model qualities. 
        
        \item We propose processing all attributes through a learnable binary classifier, which automatically adjusts the importance of different attributes and makes the final ranking. We also propose an iterative training strategy for the classifier to enable continuous refinement of ranking quality through feedback in each round.

    \end{itemize}

We evaluate GraphRank on the large-scale, small-scale, and heterophilic datasets. The results show that GraphRank significantly surpasses existing test input prioritization techniques. 

\label{sec:related}
\section{Background}
In this section, we introduce the basic knowledge of GNNs and test input prioritization techniques.

\subsection{Graph Neural Networks}
The graph is defined as \( G = (V, E) \), where \( V = \{u_1, u_2, \ldots, u_N\} \) represents the set of \( N \) nodes, and \( E \subseteq V \times V \) denotes the edge set. The features of all nodes are represented as a matrix \( \textbf{X} \in \mathbb{R}^{N \times d} \), where \( d \) is the feature dimension. Each row in the matrix \( \textbf{X} \) corresponds to the feature vector for an individual node.
The graph structure is captured by the adjacency matrix \( \textbf{A} \in \{0,1\}^{N \times N} \), which encodes the connectivity relationships between nodes. Specifically, an entry \( \textbf{A}_{ij} = 1 \) indicates an edge from the \( i^{th} \) node to the \( j^{th} \) node (i.e., \( (u_i, u_j) \in E \)), while \( \textbf{A}_{ij} = 0 \) indicates that no edge exists between those nodes (i.e., \( (u_i, u_j) \notin E \)).

In GNNs, the message-passing scheme is commonly employed under the homophily assumption. At each GNN layer, this mechanism propagates node information through the graph structure to generate effective node embeddings. In the following, we introduce a representative GNN architecture.

GNNs typically stack multiple neural network layers. For the $l_{th}$ layer, the output consists of the node embeddings $\textbf{H}^{l}$. The input to this layer includes two components: the node embeddings $\textbf{H}^{l-1}$ from the previous layer (or the initial features matrix $\textbf{X}$ if $l=1$), and the normalized adjacency matrix $\hat{\textbf{A}}$.
The definition of a stacked multi-layer GNN model is as follows:
    \begin{equation}
    \label{GNN layer 0}
    \textbf{H}^{1} = \sigma(\hat{\textbf{A}} \textbf{X} \textbf{W}^1 ),
    \end{equation}
    \begin{equation}
    \label{GNN layer n}
    \textbf{H}^l = \sigma(\hat{\textbf{A}} \textbf{H}^{l-1} \textbf{W}^l ).
    \end{equation}
Here, the matrix $\textbf{W}^l$ is the learnable parameters of the $l_{th}$ GNN layer. The function $\sigma$ denotes the activation function. The $\hat{\textbf{A}} \textbf{X}$ and $\hat{\textbf{A}} \textbf{H}^{l-1}$ are the process of message propagation and aggregation. 
The definition of the normalized adjacency matrix \( \hat{\textbf{A}} \) is different in various GNN architectures. For instance, in Graph Convolutional Networks (GCNs)~\citep{kipf2017semi}, 
$\hat{\textbf{A}} := \tilde{\textbf{D}}^{-\frac{1}{2}}\tilde{\textbf{A}}\tilde{\textbf{D}}^{-\frac{1}{2}}$,
where \( \tilde{\textbf{A}} = \textbf{A} + \textbf{I} \), and the diagonal matrix is defined as \( \tilde{\textbf{D}}_{ii} = \sum_j (\tilde{\textbf{A}}_{ij})\).

When training the $k$-layer GNNs, we first feed in $\hat{\textbf{A}}$ and $\textbf{X}$ to obtain the output $\textbf{H}^k$. Subsequently, the $\textbf{H}^k$ and node labels are fed into the loss function to train the GNN model.

\subsection{Test input prioritization}
     \textbf{Failure nodes definition:} Given a graph $G=(V, E)$ and a target node classification GNN model trained on this graph, the total number of nodes is defined as $N$, and $c$ represents the number of categories for the nodes. We denote the ground truth label for all nodes as $\textbf{Y} \in {\{1, 2, \cdots,c\}}^N$, where $\textbf{Y}_i$ represents the ground truth label for the $i_{th}$ node. The predicted labels of all nodes generated by the target GNN model are represented as $\hat{\textbf{Y}} \in {\{1, 2, \cdots,c\}}^N$. For each node, the target GNN outputs a categorical distribution $P = (p_1, p_2, \cdots, p_c)$, where $p_i$ is the probability of this node being classified as $i_{th}$ category. Therefore, the prediction label of this node is $\textbf{argmax}_i\{p_1, p_2, \cdots, p_c\}$.
    Finally, a failure node is defined when there is a mismatch between the model prediction and its ground truth label, \textit{i.e.}, the $i_{th}$ node is a failure when $\textbf{Y}_i \neq \hat{\textbf{Y}}_i$. 

    \noindent
    \textbf{Prioritization for nodes:} The objective of test input prioritization is to detect as many failures from the unlabeled inputs as possible within a given labeling budget $b$. First, we estimate the priority scores for all unlabeled nodes and rank nodes based on these scores. Second, the top $b$ nodes are then selected to constitute a selected subset, denoted as $V_s$. The objective is to maximize the number of failure nodes in $V_s$, namely maximize the number of elements in $\{u_i|\textbf{Y}_i \neq \hat{\textbf{Y}}_i, u_i \in V_s\}$. The selected nodes set $V_s$ can aid in diagnosis or repair, which facilitates the reliability and deployment of GNNs. Note that this paper focuses solely on efficient failure discovery, and the subsequent diagnosis or repair is outside our scope.

\section{Motivation of GraphRank}
\label{subsec:motivation}
    \begin{figure*}[htb]
    \centering
    \includegraphics[width=1\linewidth]{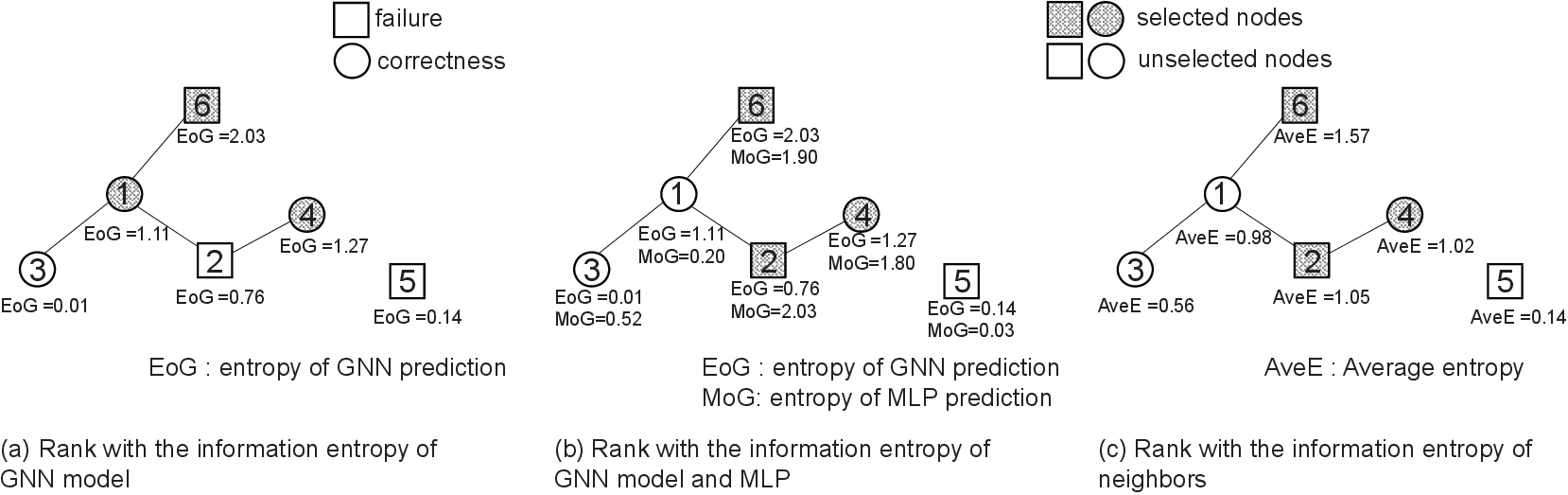}
    \caption{\textbf{A motivational example.} We set the budget for selecting nodes to be three. The shaded nodes are selected nodes, while the remaining nodes are unselected. The square nodes are failures, and the circles are correctly classified}
    \label{fig:Motivation}
    \end{figure*} 
    
In this section, we present the motivation for our method.
Figure \ref{fig:Motivation} presents a motivational example derived from realistic scenarios. Given a target GNN model and a labeling budget of three, our objective is to select unlabeled nodes that can reflect more model failures.
To facilitate understanding of different techniques, we use squares to represent nodes that fail classification, while circles denote those that can be correctly classified. 

In Fig. \ref{fig:Motivation}a, we only use the model-aware information for prioritization. Specifically, we calculate the information entropy of the categorical distributions predicted by the target GNN for each node. These values are model-aware, which reflects the uncertainty of the model. The node with a higher value is more likely to represent a failure.
We use these values as priority scores to rank all nodes and select the top three (marked with shading). However, we find that only one failure node has been selected. This phenomenon inspires us that rely solely on model-aware information may lead to suboptimal testing performance. We hypothesize that this stems from the unpredictable quality of the target model and insufficient exploration of graph structures.
When the model quality is poor, the reliability of the derived model-aware information becomes questionable. Due to the indeterminable quality of the target model, it is essential to seek additional information. Simultaneously, we also explore graph structural information to enhance the prioritization performance.

\subsection{Model-agnostic information}
Recognizing that model-aware information is heavily influenced by the target GNN model, we turn our attention to exploring model-agnostic information—data that is independent of the model. 
Specifically, we initially attempt to extract information from the graph node features matrix 
$\textbf{X}$. This feature set contributed to the node classification, provides valuable information about prioritization, and remains independent of the target GNN model. Thus, it serves as an additional source of information.

To study the efficacy of this information, we train a Multi-Layer Perceptron (MLP) solely based on graph node features to classify nodes. From the MLP's predictive distributions, we calculate the information entropy to derive model-agnostic information.  
In Fig. \ref{fig:Motivation}b, we multiply the two sources of information—one from the model-aware and the other from the model-agnostic—to generate priority scores. This approach enables the identification of an additional failure node (node 2), which was misclassified in Fig. \ref{fig:Motivation}a. This example inspires us that introducing model-agnostic information can achieve better testing performance. Therefore, we extract priority attributes from model-aware and model-agnostic, combining them for prioritization.

\subsection{Graph structure information}
In graph data, the GNNs propagate neighbor information based on the homophily assumption. Inspired by such an assumption, we aggregate priority information from neighbors, leveraging the graph structure information. 
Specifically, we aggregate the model-aware information, namely entropy, of one node and its adjacent nodes through averaging. The results are shown in Fig. \ref{fig:Motivation}c, where we also identify the failure node 2. This phenomenon exemplifies the efficacy of graph structure information, motivating us to exploit its potential.

\begin{figure*}[t]
    \centering
    \includegraphics[width=0.9\linewidth]{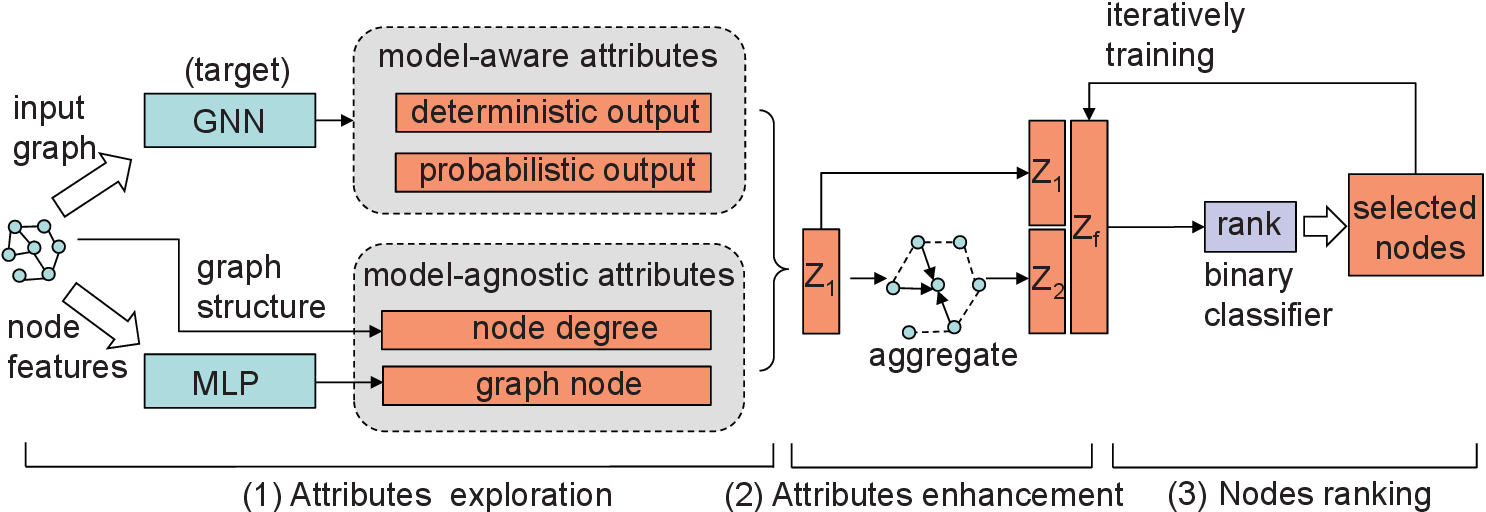}
    \caption{\textbf{The overview of GraphRank.} 1) Attribute exploration: explore both model-aware and model-agnostic attributes; 2) Attributes enhancement: enhance attributes with graph structure; 3) Nodes ranking: train a classifier for priority score estimation and ranking}
    \label{fig:overview}
\end{figure*}

\section{Method}
\label{sec:method}
      In this section, we present our test input prioritization framework, GraphRank, for GNNs.

\subsection{Overview}
    Figure \ref{fig:overview} shows the overview of GraphRank, which consists of three steps: (1) \textbf{Attributes exploration.} This step aims to explore two kinds of attributes: model-aware and model-agnostic attributes. Model-aware attributes are derived from the outputs of the target GNN model. In contrast, model-agnostic attributes are independent of the target model, which have the ability to compensate for the limitations of relying solely on model-aware attributes. (2) \textbf{Attributes enhancement.} Based on the message propagation and aggregation of GNNs and the homophily of graph data, we leverage the graph structure information to aggregate the above-explored attributes, obtaining the average values as the enhanced attributes. This step can enhance the representation of node priority. 
    (3) \textbf{Nodes ranking.} We combine attributes from steps (1, 2) and train a learnable binary classifier as a ranking model. This model predicts the priority score (i.e., the probability of being classified as 1) of each unlabeled node. We train this classifier iteratively, allowing the classifier to receive feedback and refine itself in each round.
    
    \subsection{Attributes exploration}
    \label{sec: attr_explor}
    As shown in Fig. \ref{fig:overview}, we explore two kinds of attributes: model-aware and model-agnostic attributes. 

    \subsubsection{Model-aware attributes}
    These types of attributes are directly relevant to the target GNN model. 
    Specifically, we introduce the deterministic output attributes and probabilistic output attributes. These attributes can reflect the confidence level of the model for a given node classification result.
    
    \vspace{3pt}
    \noindent
    \textbf{Deterministic output attributes.}
    The deterministic output attributes, commonly used in previous studies~\citep{li2024test3D}, are obtained directly from the original output of the target GNN model. Since the model's parameters remain unchanged after training, the predictions of the model are fixed for a certain node. Therefore, we define these attributes as deterministic.
    With the GNN model's output results, we estimate the node failure probability by measuring the dispersion in the categorical distribution.
    If the categorical distribution is more dispersed, the node is more likely to be a failure. There are various metrics that measure the dispersion of categorical distribution, including information entropy and Gini metrics~\citep{feng2020deepgini}. Given the categorical distribution of one node as $P = (p_1, p_2, \cdots, p_c)$, the calculation of entropy and Gini metrics is $-\sum_{i=1}^{c}p_i$log$(p_i)$ and $1-\sum_{i=1}^{c}p_i^2$, respectively. In GraphRank, we first compute the entropy and Gini metrics of the output categorical distribution. Then, we concatenate these two metrics and categorical distribution to form the deterministic output attributes.
    
    \vspace{3pt}
    \noindent
    \textbf{Probabilistic output attributes.}
    For a given target model, significant changes in output resulting from small perturbations indicate a high level of uncertainty, suggesting that the node is likely to fail. Building on this observation, we apply perturbations to the model to assess the failure probability of input nodes.
    Since the categorical distribution predicted by the model becomes non-deterministic after these perturbations, we denote these attributes as probabilistic output attributes.
    
    We leverage the Dropout~\citep{srivastava2014dropout} technique to estimate the probabilistic output attributes. The following describes how to extract probabilistic output attributes by Dropout. 
    In GNNs, the Dropout is applied by randomly setting some elements of the $\textbf{H}^l$ to zero with a constant rate. 
    The definition of the GNN Dropout layer is as follows:
    \begin{equation}
    \textbf{H}^{l+1} = \sigma(\hat{\textbf{A}} (\textbf{Z}^l \odot \textbf{H}^l)\textbf{W}^{l+1} ),
    \end{equation}
    where the $\textbf{Z}^l$ is a random binary mask matrix, and the proportion of 0 in $\textbf{Z}^l$ represents the dropout rate. The $\textbf{Z}^l$ takes random values during each inference calculation process, and the $\odot$ represents the element-wise product.
    To acquire multiple non-deterministic categorical distributions, we perform inference $m$ times with Dropout on the trained target GNN model, obtaining $m$ different results. Finally, we calculate the average distribution as follows:
    \begin{equation}
    AveO=softmax(\frac{1}{m}\sum_{i=1}^{m} output_i ),
    \end{equation}
    where the $output_i$ is the $i_{th}$ result and the $softmax$ is a normalization function for classification. Then, we use the information entropy of the $AveO$ as the probabilistic output attribute, which represents the uncertainty of the model for a specific node.

    \subsubsection{Model-agnostic attributes}
   The model-agnostic attributes are independent of the target GNN model. We employ these attributes to compensate for the drawback of model-aware attributes, which rely on the model's quality. Model-agnostic attributes provide priority information from the perspective of graph data, regardless of the quality of the target GNNs. In model-agnostic attributes, we explore the graph node attributes coming from the graph node features and the node degree attributes coming from the graph structure information.
   
    \vspace{3pt}
    \noindent
    \textbf{Graph node attributes.} Since GNNs classify nodes based on graph node features and structural information, the graph node features inherently contain partial classification information, which can provide valuable priority insights.
    To extract this information, we train an MLP independent of the target GNN model for the node classification task. Specifically, we first input the labeled node features $\textbf{X}_L$ and their ground truth labels $\textbf{Y}_L$ into MLP to train it. After training well, we feed the unlabeled nodes features $\textbf{X}_U$ into the MLP and obtain the output categorical distributions. 
    Finally, we calculate the information entropy of these distributions, concatenating the entropy and categorical distribution as the graph node attributes. 
    
    \vspace{3pt}
    \noindent
    \textbf{Node degree attributes.}
    One of the reasons GNNs achieve high classification performance is information propagation and aggregation. A node with more neighbors can often aggregate more information, increasing its opportunity of being classified correctly.
    Hence, the node degree, indicating the number of neighbors, reveals important information from the graph structure and can be used for prioritization. 
    We normalize the degrees of all nodes with Min-Max normalization and take the normalized degrees as the node degree attributes.

    \begin{figure}[t]
        \centering
        \includegraphics[width=8cm,height = 3.6547cm]{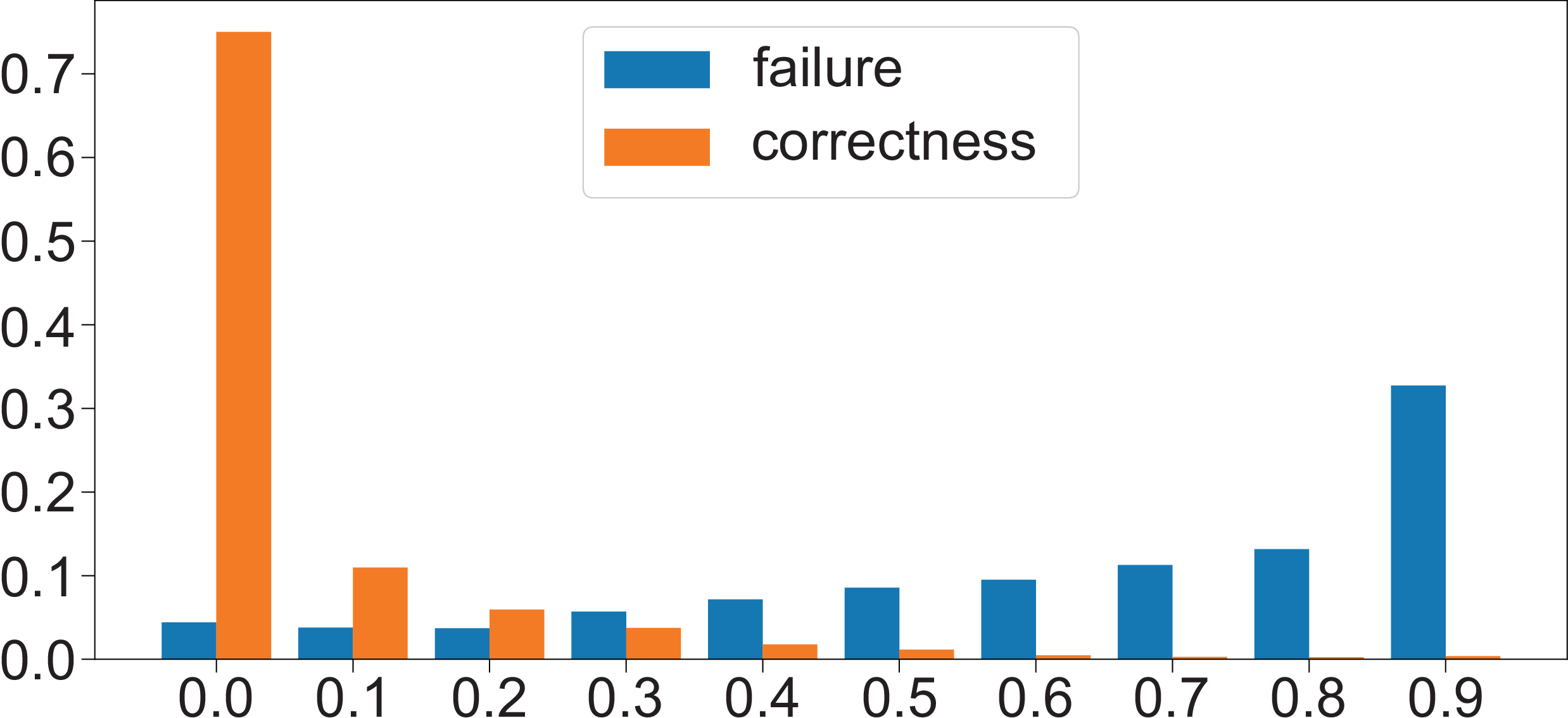}
        \caption{\textbf{The distribution of neighbor failure rates for both correctness nodes set and failure nodes set.} X-axis: the neighbor failure rates. Y-axis: the proportion of nodes corresponding to different neighbor failure rates}
        \label{nei fail rate}
    \end{figure}
    
    \subsection{Attributes enhancement}
    After extracting all the above attributes, we concatenate them as $\textbf{Z}_1 \in R^{N \times d_1}$, where the $N$ is the number of nodes and $d_1$ is the dimension of an attribute vector.

According to the homophily hypothesis, if the majority of a node's neighbors are failures, that node is more likely to fail, and vice versa. We validate this hypothesis in Fig. \ref{nei fail rate}.
In this figure, we categorize nodes into two groups: failure nodes and correct nodes, based on their classification results. We then calculate the neighbor failure rate for each node, which is defined as the number of failing neighbors divided by the total number of neighbors. We separately plot the distribution of neighbor failure rates for both correctness nodes set and failure nodes set.
As illustrated in Fig. \ref{nei fail rate}, failed nodes exhibit a relatively high neighbor failure rate, while correct nodes show the opposite trend. This observation supports the notion that most nodes display homophily.
Motivated by this finding, we propose aggregating the attributes of neighboring nodes to enhance overall attributes. Specifically, we normalize the adjacency matrix \( \textbf{A} \) to obtain \( \textbf{A}' \), where 
$
\textbf{A}'_{ij} = \frac{(\textbf{A} + \textbf{I})_{ij}}{\sum_{k=1}^N \textbf{A}_{ik} + 1}.
$
We then aggregate the neighboring attributes using \( \textbf{A}' \), denoting this aggregated result as 
$
\textbf{Z}_2 = \textbf{A}'\textbf{Z}_1.
$
Consequently, \( \textbf{Z}_2 \) represents the enhanced attributes that incorporate graph structure information.
    
    However, we also observe that some failure nodes have low neighbor failure rates, indicating that they are non-homophilic. Handling these non-homophilic nodes solely based on $\textbf{Z}_2$ may present challenges. Therefore, we concatenate the $\textbf{Z}_1$ and $\textbf{Z}_2$ to form the final attributes $\textbf{Z}_f$. The retained $\textbf{Z}_1$, which contains some non-structural information (\textit{i.e.}, graph node attributes), is necessary for these nodes to be identified correctly. In this step, we retain sufficient information, allowing the ranking model in the subsequent step to automatically learn the importance of attributes for both homophilic and non-homophilic nodes, thereby ensuring effective handling of both types of nodes.

    \subsection{Nodes ranking}
    Having obtained the final attributes $\textbf{Z}_f$, we train a binary classifier as the ranking model to predict the priority scores. The training process of the classifier is described as follows.
    \subsubsection{Training set construction}
    \label{sec: construct}
    Given the labeled nodes set $V_L$ and unlabeled nodes set $V_U$, we denote the final attributes on the labeled nodes set and unlabeled nodes set as $\textbf{Z}_L$ and $\textbf{Z}_U$, respectively. On the labeled nodes set, we can identify whether these nodes are misclassified and assign the binary labels to nodes based on the classification results. If the target model misclassifies a node, we label it as 1, and if not, we label it as 0. These binary labels on $V_L$ are defined as $\textbf{C}_L$. We take the ($\textbf{Z}_L$, $\textbf{C}_L$) as the training set of the binary classifier, where $\textbf{Z}_L$ is the inputs and $\textbf{C}_L$ is the training labels.
    
    In practical application scenarios, it is not reasonable to utilize the original training nodes set of the target GNN model to construct ($\textbf{Z}_L$, $\textbf{C}_L$). Since the GNN is trained on the training nodes set, its accuracy on these nodes can be higher than the rest of the unseen nodes. Directly employing the original training nodes set to construct the ($\textbf{Z}_L$, $\textbf{C}_L$) and train the binary classifier may lead to poor generalization. Due to the difference in GNN's accuracy between the training nodes set and the unseen nodes set, the distribution of binary labels is different. Therefore, the additional unseen labeled data is required to construct ($\textbf{Z}_L$, $\textbf{C}_L$). In general, the accessibility of these data is not an obstacle. Engineers often have a validation set to perform a preliminary model quality evaluation. This validation set can be used to construct the training set of the binary classifier. 
    
    \subsubsection{Training the ranking model}
     We train a binary classifier as the ranking model with the constructed training set ($\textbf{Z}_L$, $\textbf{C}_L$). The output of the binary classifier is a continuous value ranging from 0 to 1, representing the probability of the input being a failure. If this value exceeds 0.5, the classification result of the binary classifier is 1; otherwise, the result is 0. GraphRank only requires the continuous values that represent the likelihood of a node being misclassified. We serve these values as priority scores and utilize them for subsequent ranking. 

    \subsubsection{Iteratively training}

To enhance the performance of the binary classifier, we employ an iterative training strategy, as outlined in Algorithm \ref{ite train}. We begin by dividing the labeling budget, denoted as \( b \), into multiple equal portions, each referred to as the round budget \( b' \).
In each round, we train the binary classifier using the current training set \( (\textbf{Z}_L, \textbf{C}_L) \). With this classifier, we estimate the priority scores for unlabeled nodes and annotate the top \( b' \) nodes based on these scores. Once the top \( b' \) ranked nodes are labeled, they are incorporated into the current training set \( (\textbf{Z}_L, \textbf{C}_L) \) to create a new training set for the subsequent round of training.
This iterative process continues until the labeling budget is fully exhausted.

    \begin{algorithm}
        \renewcommand{\algorithmicrequire}{\textbf{Input:}}
        \renewcommand{\algorithmicensure}{\textbf{Output:}}
        \caption{Training binary classifier iteratively}
        \label{ite train}
        \begin{algorithmic}[1]
            \Require 
            \Statex $\textbf{Z}_L$: final attributes on labeled nodes set 
            \Statex $\textbf{Z}_U$: final attributes on unlabeled nodes set 
            \Statex $\textbf{C}_L$: binary labels 
            \Statex $b$: labeling budget  
            \Statex $b'$: round budget  
            
            \Ensure 
            \Statex  $\textbf{Z}_{select}$: Selected unlabeled nodes set 
            
            \Statex \hspace*{-\leftmargin} ${\textbf{Function:}}$ 
            \State $\textbf{Z}_{select}$ = $\varnothing$    
            \While{$b>0$}
            \State {Classifier} = {train}$(\textbf{Z}_L, \textbf{C}_L)$       { // Train classifier}
            \State $\textbf{scores}$ = {Classifier}$(\textbf{Z}_U)$           {  // Calculate the priority scores of unlabeled nodes}
            \State $\textbf{Z}_U$ = {sort}($\textbf{Z}_U$, $\textbf{scores}$)    {  // Sort the $\textbf{Z}_U$ by priority scores}
            \State $\textbf{Z}_{new}$ = {top}($\textbf{Z}_U$, {min}$\{b', b\}$)  {  // Select unlabeled nodes}
            \State $\textbf{C}_{new}$ = {annotate}($\textbf{Z}_{new}$)  
            \State $b$ = $b-${min}$\{b', b\}$
            \State $\textbf{Z}_L$ = $\textbf{Z}_L \cup \textbf{Z}_{new}$, $\textbf{C}_L$ = $\textbf{C}_L \cup \textbf{C}_{new}$, $\textbf{Z}_U$ = $\textbf{Z}_U - \textbf{Z}_{new}$ 
            \State $\textbf{Z}_{select}$ = $\textbf{Z}_{select} \cup \textbf{Z}_{new}$  
            \EndWhile
            
            \State \Return $\textbf{Z}_{select}$
        \end{algorithmic}
    \end{algorithm}

\section{Experiment design}
\subsection{Research questions}
We investigate the following research questions.

    \subsubsection{RQ1: What is the performance of GraphRank?}
    In this research question, we assess the testing performance of GraphRank by comparing it with baseline methods on large-scale graph datasets, employing the established test input prioritization metrics, TRC and ATRC~\citep{li2021testrank}. Furthermore, we evaluate the significance of the improvements achieved by our method over baseline methods using statistical significance metrics. These evaluations collectively enable us to determine the effectiveness of GraphRank in addressing the test input prioritization task for GNNs.

    \noindent
    \textbf{Experimental design:} We leverage three large-scale graph datasets and four scalable GNN models, leading to twelve combinations. In these combinations, we evaluate the testing performance of GraphRank and nine baseline methods using the ATRC metric. We also plot the TRC curves for each method across different budget sets, with each curve representing the average TRC value aggregated over twelve combinations. The details of the compared methods can be found in Section \ref{sec:baseline}. 
    Moreover, since randomness exists in machine learning, we conducted four repetitions of the experiments to reduce the impact of randomness. The following results are the averages obtained from the four repeated experiments. 
    
    We leverage the $p$-value and effect size to assess whether GraphRank significantly outperforms the baseline methods. The $p$-value and effect size are metrics for determining the statistical significance of differences between the results of two groups. For these metrics, we adopt the calculation methods employed in NodeRank, namely the Mann-Whitney U test~\citep{mcknight2010mann} and Cohen's $d$~\citep{cohen2016power}. We calculated the $p$-value and effect size based on the results from GraphRank and the comparison methods. A $p$-value $<$ 0.05 and an effect size $>$ 0.8 generally indicate statistical significance.

    \subsubsection{RQ2: How do the specific attributes work in GraphRank?}
    
    In GNN testing, nodes are classified as either correct or failure, and the priority attributes are designed to distinguish these two categories. To investigate the effectiveness of each attribute, we conduct an individual analysis of the four attributes used in GraphRank. In this research question, we analyze the distribution of each attribute across the two classes of nodes to assess its ability to identify correctness and failure nodes.
    
    \noindent
    \textbf{Experimental design:} 
    To achieve our objective, we prioritize unlabeled test nodes using each attribute independently.
    For deterministic and probabilistic output attributes, we calculate their information entropy as a priority metric. 
    For graph node and node degree attributes, we leverage the entropy of MLP output and degrees of nodes as the priority metrics, respectively. 
    After extracting the priority scores from these four attributes, we plot their priority score distributions on the failure nodes set and correctness nodes set. To compare the gaps between these individual metrics and the final priority scores, we also show the distribution of the final priority scores estimated by GraphRank.

    \subsubsection{RQ3: Are all components useful in GraphRank?}
    GraphRank primarily consists of four components: model-aware attributes, model-agnostic attributes, attributes enhancement, and iterative training. To measure the contribution of each component, we conduct the corresponding ablation studies to analyze their effectiveness. 

    \noindent
    \textbf{Experimental design:} 
    In GraphRank, we consider the model-aware attributes a key component since they are obtained from the GNNs and directly reflect the probability of failure. Then, the model-agnostic attributes are assumed to be the next most important. Finally, attributes enhancement and iterative training are considered the third and fourth components to be incorporated in an ablation study, respectively. The steps for investigating the contributions of each component are as follows:
    
    \begin{enumerate}
        \item We evaluate the testing performance of GraphRank with only model-aware attributes (GraphRank$_{AW}$).
        
        \item We evaluate the testing performance of GraphRank with model-aware attributes and model-agnostic attributes (GraphRank$_{AW+AG}$).
        
        \item We evaluate the testing performance of GraphRank with two types of attributes and attributes enhancement (GraphRank$_{AW+AG+EN}$).
        
        \item We evaluate the testing performance of GraphRank with all components (GraphRank$_{Complete}$).
    \end{enumerate}
    
    On the other hand, we can easily incorporate other attributes since GraphRank is a flexible framework that considers both model-aware and model-agnostic attributes. We note that the mutation attributes extracted from GraphPrior and NodeRank are model-aware, which can also be included in our framework. To validate the superiority of our framework, we replace the model-aware attributes in GraphRank with the mutation attributes derived from GraphPrior and NodeRank, respectively.

    \subsubsection{RQ4: Could the failures detected by the testing methods improve the performance of GNNs?}
    
    Following the testing phase, the selected test inputs are leveraged for downstream model repair tasks. To verify the effectiveness of detected failure nodes, we conduct the model repair experiment. Please note that the purpose of this experiment is only to evaluate the repair capability of different test input prioritization techniques, and model repair is beyond the scope of this paper.

    \noindent
    \textbf{Experimental design:}
    We employ a simple model retraining method for this experiment.
    First, we set the labeling budget as one-fifth of the training set, leveraging the test input prioritization methods to select nodes from the test set. These chosen nodes are incorporated into the training set, and we retrain the GNNs on the new augmented training set. Since some test nodes are included in the training set, measuring the performance improvement on the test set is inappropriate. Therefore, we leverage the performance improvement on the hold-out validation set to assess the impact of the selected nodes on model repair.

    \subsubsection{RQ5: What is the time overhead of GraphRank?}
    In the testing framework we propose, GraphRank, the processes of attributes extraction and ranking model training result in a non-negligible time overhead. Similarly, existing mutation-based methods require training additional mutant models, which also involves substantial computational cost. In this research question, we evaluate the runtime efficiency of GraphRank and the baseline methods.

    \noindent
    \textbf{Experimental design:}
    In this research question, we compare the time overhead of GraphRank with other baseline methods.
    In all compared methods, Dropout, GraphPrior, and NodeRank require additional model training or inference, whereas the other methods operate without such time overhead. So we consider other compared methods as zero-cost. However, these zero-cost methods often demonstrate limited performance, rendering them inadequate for the effective testing of GNNs. 
    Consequently, we focus on the time overhead of Dropout, GraphPrior, NodeRank, and GraphRank. We assess their computational costs on the EnGCN model with the experimental platform described in Section \ref{sec: implement}. 

    In addition, since GraphRank incorporates iterative training, the number of iterations also becomes a key factor influencing time overhead. We investigate the performance of GraphRank under varying iteration counts.
    
    \subsubsection{RQ6: What is the adaptability of GraphRank to other types of graph datasets?}

    In the previous research question, we conducted experiments only on the large-scale datasets. To validate the adaptability capability of GraphRank on other types of datasets, we evaluate the performance of GraphRank on small-scale and heterophilic graph datasets.
            
    \noindent
    \textbf{Experimental design:}
    For the small-scale dataset, we leverage three graph datasets (Cora, CiteSeer, and PubMed) and three regular GNN models (GAT, GCN, and TAGCN). On these models and datasets, we evaluate the testing performance of GraphRank and nine baseline methods, using ATRC as the metric. For the heterophilic graphs, we leverage two graph datasets (Roman-empire and Amazon-ratings) and two GNN models (GAT and Dir-GNN), where the Dir-GNN is the heterophilic GNN model. We also evaluate all testing methods based on the ATRC metric.
    
    \subsubsection{RQ7: How does the selection of the binary classifiers impact the effectiveness of GraphRank?}
    
    In GraphRank, we leverage the binary classifier as the ranking model. The choice of classifier may influence the performance of GraphRank. To investigate this impact, we evaluate GraphRank using various classifiers, including three based on decision trees and two based on neural networks.
    
    \noindent
    \textbf{Experimental design:}
    We select a variety of classifiers, including XGBoost~\citep{chen2016xgboost}, LightGBM~\citep{ke2017lightgbm}, Random Forest~\citep{breiman2001random}, MLP, and ClusterGCN classifier. Please note that to reduce the influence of other factors, we removed iterative training in this experiment. We evaluate the testing performance of GraphRank under different classifiers in terms of ATRC.

\subsection{Evaluation criteria}
The Average Percentage of Fault Detection (APFD)~\citep{yoo2012regression} is one of the criteria leveraged to evaluate the overall performance of test input prioritization techniques. However, such criteria are independent of the labeling budget and cannot reflect the performance of specific methods under a given budget. Since the performance of GraphRank is related to budget setting due to the iterative training process, APFD fails to evaluate GraphRank. Therefore, we explore other reasonable criteria that consider the budget setting.

We adopt the new criteria, Test Relative Coverage (TRC) and Average Test Relative Coverage (ATRC)~\citep{li2021testrank}, to evaluate the test input prioritization techniques. These criteria take the budget setting into account. With the given labeling budget $b$, TRC measures the gap between the current test input prioritization technique and an ideal technique, which is defined as follows:

\begin{equation}
TRC = \frac{DF}{min\{b, TF\} },
\end{equation}

where $b$ is the labeling budget, $TF$ represents the number of total failures in the unlabeled nodes set, and $DF$ represents the number of detected failures with budget $b$. 

ATRC calculates the average TRC values across various budget settings. Consistent with prior work~\citep{li2021testrank}, we restrict the budget below the number of total failures. The ATRC is defined as follows:
\begin{equation}
ATRC =  \frac{1}{N} \sum\limits_{i=1}^{N}{TRC_i},
\end{equation}
where $TRC_i$ represents the $TRC$ value at budget $b_i$, and all $b_i$ values are lower than $TF$. To verify the overall quality of the test input prioritization techniques, we set the budget ranging from one-tenth of the total failures to all total failures in the experiment. ATRC considers different budgets and comprehensively evaluates the performance of GraphRank. 

\subsection{Baseline methods}
\label{sec:baseline}
We consider nine baselines, including random selection, six test input prioritization techniques designed for the DNNs and two techniques designed for the GNNs. For the categorical distribution of a node predicted by the target GNN model, we define it as $P = (p_1, p_2, \cdots, p_c)$, where $c$ is the number of categories. 

\textbf{Random}~\citep{elbaum2002test}. This method ranks and selects the unlabeled nodes randomly.

\textbf{Entropy}~\citep{weiss2022simple}. This method calculates the information entropy of categorical distribution as a priority score. The formula of this metric is $-\sum_{i=1}^{c}p_i$log$(p_i)$.

\textbf{DeepGini}~\citep{feng2020deepgini}. This method calculates the Gini metric of categorical distribution as a priority score. The formula of this metric is $1-\sum_{i=1}^{c}p_i^2$.

\textbf{Dropout}~\citep{srivastava2014dropout}. 
For each unlabeled node, this method employs the Dropout technique to generate multiple distinct categorical distributions. It then averages these distributions and calculates the information entropy of the averaged distribution to derive a priority score.

\textbf{Margin}~\citep{wang2014new}. This method calculates the Margin metric of categorical distribution as a priority score. The formula of this metric is $p_i-p_j$, where $p_i$ is the second maximum value in the distribution and $p_j$ is the maximum value.

\textbf{DATIS}~\citep{li2024distance}. For each test input, this method finds its nearest neighbor samples to generate a neighbor support vector. Specifically, given the labels of neighbor samples, DATIS calculates their contributions to the central node for each corresponding label. Through summing and normalizing these contributions, DATIS generates the support vector for each test input. The vector, whose dimensionality corresponds to the number of classes, represents the extent to which neighboring nodes support the prediction of the central node. Finally, DATIS estimates the uncertainty of one test input by combining its support vector and categorical distribution.

\textbf{NNS}~\citep{bao2023defense}. This method is inspired by the label smoothing technique, which smooths the categorical distribution for each test input. Specifically, this metric calculates a weighted average of the categorical distributions of the neighbor and the test input. Then, it leverages existing metrics, such as Gini, to calculate the uncertainty of the averaged categorical distribution, which is then used as an indicator of failure probability for prioritizing test inputs.

\textbf{GraphPrior}~\citep{dang2023graphprior}. This method designs various mutation rules for GNNs, measuring the failure probability by analyzing whether the mutants are killed. The ``killed'' means that the predictions of a mutant differ from the original GNN model. Given model's mutants $\{M_1, M_2,\cdots , M_n\}$, they define the mutation attribute as $\textbf{K}\in \{0,1\}^n$. If the $M_i$ is killed, the $\textbf{K}_i = 1$, otherwise, the $\textbf{K}_i = 0$.
Finally, GraphPrior trains a ranking model with the mutation attributes to estimate the priority score from each node.

\textbf{NodeRank}~\citep{li2024test}. This method also designs mutation rules specifically for GNNs. However, NodeRank trains multiple ranking models with different architectures, employing the ensemble approach~\citep{mienye2022survey, divina2018stacking} to integrate multiple priority scores for prioritization.

\subsection{Datasets and models} 
\subsubsection{Large-scale level}
We evaluate our method with three large-scale graph datasets of the node classification tasks. 
The Ogbn-products~\citep{hu2020open} dataset is an Amazon product co-purchasing network, where the nodes represent products. 
The Reddit~\citep{hamilton2017inductive} dataset represents the social network of Reddit, where the nodes are posts in different communities.
Flickr~\citep{zenggraphsaint} dataset containing descriptions and common properties of images, where each node represents an image. 
To mimic realistic scenarios, we swap the original training and test sets on the Reddit and Flickr datasets. Swapping training and test sets have negligible impacts on the performance of the target GNNs. The average changes in the accuracies of GNNs on the Reddit and Flickr datasets are only 1.37\% and 2.30\%, respectively. The details of these datasets are shown in Table \ref{tab:large_data}.
 
We adopt four representative scalable GNN models that can be used to handle large-scale Graphs: EnGCN~\citep{duancomprehensive}, SIGN~\citep{frasca2020sign}, GraphSAGE~\citep{hamilton2017inductive}, and ClusterGCN~\citep{chiang2019cluster}. 
These models utilize sampling-based or decoupling-based methods to reduce GPU memory usage, enabling their suitability for large-scale graph datasets. Table \ref{tab:lar_model_acc} presents the accuracies of these models on different datasets. 

\begin{table}
\caption{The details of the large-scale graph datasets}
\centering
\label{tab:large_data}
\begin{tabular}{c c c c c}
  \toprule
  \textbf{Dataset} & \textbf{Class} & \textbf{Nodes} & \textbf{Edges} & \textbf{Sparsity}\\
  \midrule
  Ogbn-products & 47 & 2449029 & 61859140 & 1.03$\times10^{-5}$  \\
  Reddit & 41 & 232965 & 57307 & 1.06$\times10^{-5}$   \\
  Flickr & 7 & 89250 & 449878 & 5.65$\times10^{-5}$   \\
  \bottomrule
\end{tabular}
\end{table}

\begin{table}[t]
\centering
    \caption{The accuracy of the scalable target GNNs on the test set ($\%$)}
    \label{tab:lar_model_acc}
    \begin{tabular}{c c c c c}
        \toprule
        \textbf{Dataset} & \textbf{EnGCN} & \textbf{SIGN}  & \textbf{GraphSAGE}  & \textbf{ClusterGCN}  \\ 
         \midrule
         Ogbn-products  & 75.34 & 75.06 & 78.21 & 76.91 \\
         Reddit         & 93.84 & 95.24 & 95.41 & 94.82 \\
         Flickr         & 51.22 & 51.03 & 52.28 & 48.90 \\
         \bottomrule
    \end{tabular}
\end{table}

\subsubsection{Small-scale level}
To validate the generalization of GraphRank, we conduct experiments on three small-scale graph datasets. The Cora~\citep{yang2016revisiting}, CiteSeer~\citep{yang2016revisiting}, and PubMed~\citep{yang2016revisiting} datasets are citation networks, where the nodes represent publications and edges denote citation relationships. We split the dataset using a 4:3:3 ratio for training, validation, and test. Dataset statistics are presented in the top half of Table \ref{tab:other_data}. For the selection of models, we use the three regular GNN models: GAT~\citep{velivckovic2018graph}, GCN~\citep{kipf2017semi}, and TAGCN~\citep{du2017topology}. Although these models are unsuitable for large-scale graphs, they can efficiently handle small-scale datasets.

\subsubsection{Heterophilic graphs}
We investigate the performance of GraphRank on heterophilic graphs. Heterophily, the opposite of homophily, characterizes graphs where adjacent nodes are less likely to share the same label. In contrast, homophilic graphs exhibit a higher probability that connected nodes share the same label.
Specifically, we use two heterophilic graph datasets: Roman-empire~\citep{platonovcritical} and Amazon-ratings~\citep{platonovcritical}. 
The Roman-empire is derived from the Roman Empire article on English Wikipedia, where each node represents one word in the text, and edges connect words with a following or dependency relationship. The label of a node is its syntactic role.
The Amazon-ratings dataset is derived from the Amazon product co-purchasing network, where the nodes are products and edges connect products frequently bought together. The label of each node represents the average rating given by a purchaser to a product, categorized into five levels.
Based on the dataset's label and edge descriptions, it is evident that there is no explicit correlation between labels and edges. Consequently, connected nodes may not share the same label, thereby introducing heterophily into the network.
We use official node segmentation, and dataset details are presented in the lower half of Table \ref{tab:other_data}.
We utilize two GNN models, GAT and Dir-GNN~\citep{rossi2024edge}. The GAT is a regular model, while the Dir-GNN is a heterophilic GNN model.

\begin{table}
\caption{The details of the small-scale and heterophilic graph datasets}
\centering
\label{tab:other_data}
\begin{tabular}{c c c c c}
  \toprule
  \textbf{Dataset} & \textbf{Class} & \textbf{Nodes} & \textbf{Edges} & \textbf{Sparsity}\\
  \midrule
  Cora & 7 & 2708 & 5283 & 7.20$\times10^{-4}$  \\
  CiteSeer & 6 & 3327 & 4552 & 4.11$\times10^{-4}$   \\
  PubMed & 3 & 19717 & 44324 & 1.14$\times10^{-4}$   \\ \midrule
  Roman-empire & 18 & 22662 & 32927 & 6.10$\times10^{-5}$  \\
  Amazon-ratings & 5 & 24492 & 93050 & 1.55$\times10^{-4}$    \\
  \bottomrule
\end{tabular}
\end{table}

\subsection{Implementation}
\label{sec: implement}
For the implementation of GraphRank, we iteratively train the binary classifier ten times and consume ten percent of the given labeling budget to annotate selected nodes each round. Concerning the selection of the binary classifier, we use the XGBoost~\citep{chen2016xgboost} as the binary classifier. The implementation of XGBoost is based on the \textit{xgboost} package in \textit{Python}.
The implementation of the Dropout technique is achieved by performing inference ten times with the 0.5 Dropout rate for each GNN. For GraphPrior and NodeRank, the mutation rules require GNN retraining, making them unaffordable to implement on large-scale graph datasets. Considering the training costs, we set the number of mutants to six, ensuring the time overhead is acceptable.

We implement GraphRank and baselines in Python 3.10 based on Pytorch 2.3.0, PyG 2.5.3, and XGBoost 2.0.3.
The experiment is conducted on a high-performance GPU server, running on a 2.6 GHz Intel Xeon Glod 6240 CPU and an NVIDIA GeForce RTX 3090 GPU.

\section{Experiment results}
For each research question, we present the results and analysis, subsequently answering the question. All performance comparisons in this section are based on absolute increase in percentage points.

\subsection{RQ1: Performance of GraphRank}

\begin{figure*}[t]
    \centering
    \includegraphics[width=1\linewidth]{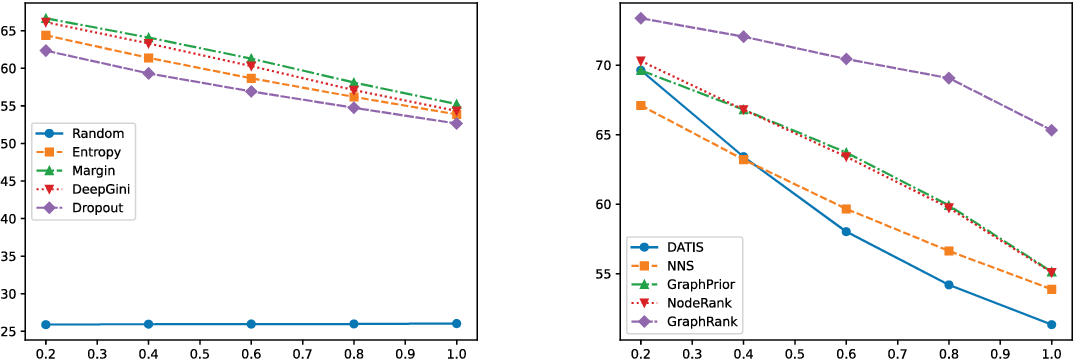}
    \caption{The TRC curves of GraphRank and the baseline methods. X-axis: The ratio of budget allocation to the total number of failures. Y-axis: TRC value (\%)}
    \label{fig:TRC_curve}
\end{figure*}

\textbf{Results:} The experiment results of RQ1 are presented in Table \ref{tab:ATRC_EX}, Table \ref{tab:pv}, and Fig. \ref{fig:TRC_curve}. We highlight the best results in bold. 
Table \ref{tab:ATRC_EX} presents the performance of baselines and GraphRank based on the ATRC metric. We provide the standard deviation of the GraphRank results.
We have several observations from this table. First, generally speaking, GraphRank outperforms all the compared baseline methods. In the Ogbn-products dataset, GraphRank demonstrates the most significant improvement, i.e., 4.31\% improvement on average. This indicates that GraphRank is more suitable for detecting GNN failures. 
Second, we note that the performance of “Random” is related to the accuracy of a model. The higher model accuracy leads to poorer performance of the “Random”. For instance, the accuracies of four GNNs in the Reddit dataset are 93.84\%, 95.24\%, 95.41\%, and 94.82\%. However, the corresponding ATRC values are 6.39\%, 4.69\%, 4.40\%, and 5.13\%, respectively. This shows the necessity of developing test input prioritization techniques, especially for high-performance models. 
Third, on the Reddit dataset, the performance of GraphRank is not as significant as that in the other two datasets. According to our analysis, this phenomenon can be attributed to the fact that the GNNs exhibit the highest accuracies on the Reddit dataset. Remember that we train a binary classifier that automatically determines the importance of model-aware and model-agnostic attributes for ranking. For high-accuracy models, the quality of model-aware attributes is high. Then, the classifier learns to pay more attention to these attributes and diminishes the importance of the model-agnostic ones, leading to limited performance improvement over model-aware-based methods. However, even if GraphRank does not show the best performance on every model, it still is the best method as it has eleven best-performed cases, while the others have at most one case.

Figure \ref{fig:TRC_curve} presents the TRC curves for each method, where the X-axis represents the ratio of budget allocation to the total number of failures, and the Y-axis indicates the TRC value. It can be observed that the TRC curve for the Random method is nearly flat, indicating that its TRC values are independent of the budget and depend solely on the proportion of failure nodes. On the other hand, we observe that the TRC values of other methods decrease as the budget increases, indicating that the efficiency of the testing methods declines when the budget becomes large. Finally, GraphRank consistently achieves higher TRC values than other methods, validating its superior efficiency.

Table \ref{tab:pv} presents the $p$-value and effect size for GraphRank compared to other test input prioritization methods. In most cases, the improvements of GraphRank are statistically significant. Although the improvement over GraphPrior and NodeRank is less significant, GraphRank achieves an average performance increase of 5.02\% and 3.06\% with lower computational overhead. Therefore, we conclude that the improvement in GraphRank is meaningful.

\noindent
\textbf{Answer to RQ1:} GraphRank consistently outperforms all compared methods, including GraphPrior, NodeRank, and others. From the results of statistical significance, GraphRank exhibits significant performance improvements in most of the comparisons.

\begin{table}[]
\centering
\caption{Comparing GraphRank with baseline methods using ATRC ($\%$). We average the results from multiple runs of experiments}
\label{tab:ATRC_EX}
\begin{tabular}{cl|cccc} \toprule
\textbf{Dataset}                                                           & \textbf{Method}    & \textbf{EnGCN} & \textbf{SIGN}  & \textbf{GraphSAGE} & \textbf{ClusterGCN} \\ \midrule
\multirow{10}{*}{\begin{tabular}[c]{@{}c@{}}Ogbn-\\ products\end{tabular}} & Random             & 24.77          & 24.91          & 21.85              & 22.99               \\
                                                                           & Entropy            & 63.60          & 72.65          & 60.32              & 62.67               \\
                                                                           & Margin             & 64.98          & 71.07          & 63.91              & 67.06               \\
                                                                           & DeepGini           & 63.90          & 73.01          & 62.31              & 65.52               \\
                                                                           & Dropout            & 63.87          & 69.34          & 59.95              & 62.37               \\
                                                                           & DATIS              & 50.25          & 76.12          & 69.00              & 69.04               \\
                                                                           & NNS                & 65.22          & 77.72          & 65.72              & 66.31               \\
                                                                           & GraphPrior         & 62.24          & 75.41          & 70.20              & 69.21               \\
                                                                           & NodeRank           & 72.43          & 75.48          & 70.16              & 70.34               \\
                                                                           & \textbf{GraphRank} & \textbf{82.64 $\pm$ 1.01} & \textbf{81.47 $\pm$ 0.91} & \textbf{72.06 $\pm$ 0.48}     & \textbf{71.74 $\pm$ 0.21}      \\ \midrule
\multirow{10}{*}{Reddit}                                                   & Random             & 6.39           & 4.69           & 4.40               & 5.13                \\
                                                                           & Entropy            & 62.65          & 62.34          & 52.23              & 47.52               \\
                                                                           & Margin             & 61.97          & 61.03          & 56.37              & 56.47               \\
                                                                           & DeepGini           & 64.57          & 63.72          & 55.74              & 52.24               \\
                                                                           & Dropout            & 64.34          & 59.76          & 51.39              & 46.29               \\
                                                                           & DATIS              & 51.57          & 57.43          & 48.50              & 56.82               \\
                                                                           & NNS                & 57.74          & 59.66          & 48.75              & 49.57               \\
                                                                           & GraphPrior         & 65.12          & \textbf{69.17} & 53.74              & 59.99               \\
                                                                           & NodeRank           & 66.69          & 68.45          & 56.29              & 58.34               \\
                                                                           & GraphRank          & \textbf{69.76 $\pm$ 0.12} & 62.59 $\pm$ 0.40          & \textbf{57.08 $\pm$ 0.11}     & \textbf{66.13 $\pm$ 0.40}      \\ \midrule
\multirow{10}{*}{Flickr}                                                   & Random             & 48.95          & 47.86          & 47.76              & 50.50               \\
                                                                           & Entropy            & 60.21          & 56.95          & 60.48              & 56.91               \\
                                                                           & Margin             & 62.63          & 60.74          & 61.44              & 61.61               \\
                                                                           & DeepGini           & 61.30          & 58.37          & 61.85              & 58.99               \\
                                                                           & Dropout            & 56.52          & 56.84          & 60.59              & 56.23               \\
                                                                           & DATIS              & 67.15          & 54.42          & 56.69              & 57.47               \\
                                                                           & NNS                & 61.22          & 51.45          & 60.46              & 58.70               \\
                                                                           & GraphPrior         & 63.45          & 63.67          & 52.99              & 63.94               \\
                                                                           & NodeRank           & 66.74          & 62.48          & 61.89              & 63.37               \\
                                                                           & GraphRank          & \textbf{70.90 $\pm$ 0.20} & \textbf{65.17 $\pm$ 0.12} & \textbf{64.57 $\pm$ 0.13}     & \textbf{65.31 $\pm$ 0.21}      \\ \bottomrule
\end{tabular}
\end{table}

\begin{table}[]
\centering
\caption{Significance analysis comparing GraphRank and baseline methods (in terms of $p$-value and effective size)}
\label{tab:pv}
\begin{subtable}{}
    \begin{tabular}{ccccc} \toprule
    \textbf{Significance} & \textbf{\begin{tabular}[c]{@{}c@{}}GraphRank \\ vs Entropy\end{tabular}} & \textbf{\begin{tabular}[c]{@{}c@{}}GraphRank \\ vs Margin\end{tabular}} & \textbf{\begin{tabular}[c]{@{}c@{}}GraphRank \\ vs DeepGini\end{tabular}} & \textbf{\begin{tabular}[c]{@{}c@{}}GraphRank \\ vs Dropout\end{tabular}} \\ \midrule
$p$-value & 0.0029 & 0.0086 & 0.0094 & 0.0009 \\
effect size & 1.4109 & 1.1673 & 1.1926 & 1.5631 \\ \bottomrule                                            
    \end{tabular}
\end{subtable}
\begin{subtable}{}
    \begin{tabular}{ccccc} \toprule
    \textbf{Significance} & \textbf{\begin{tabular}[c]{@{}c@{}}GraphRank \\ vs DATIS\end{tabular}} & \textbf{\begin{tabular}[c]{@{}c@{}}GraphRank \\ vs NNS\end{tabular}} & \textbf{\begin{tabular}[c]{@{}c@{}}GraphRank \\ vs GraphPrior\end{tabular}} & \textbf{\begin{tabular}[c]{@{}c@{}}GraphRank \\ vs NodeRank\end{tabular}} \\ \midrule
$p$-value & 0.0166 & 0.0166 & 0.0690 & 0.4025 \\
effect size & 1.2383 & 1.1957 & 0.7519 & 0.4830 \\ \bottomrule                                            
    \end{tabular}
\end{subtable}
\end{table}

\subsection{RQ2: In-depth analysis of specific attributes}

\begin{figure*}[t]
    \centering
    \includegraphics[width=1\linewidth]{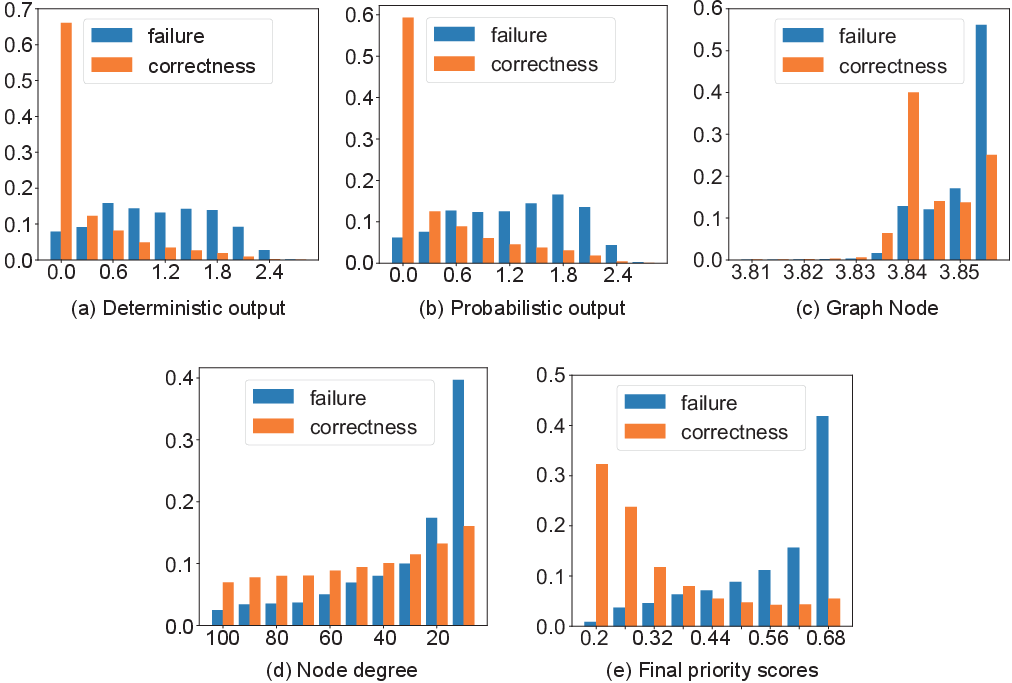}
    \caption{The distribution of each priority metric on failure nodes and correctness nodes. X-axis: the corresponding values of each priority metric. Y-axis: the proportion of nodes (\%)}
    \label{tab:metric_distribution}
\end{figure*}

\textbf{Results:} 
The experiment results of RQ2 are presented in Fig. \ref{tab:metric_distribution}, where the X-axis corresponds to the distribution of priority scores, and the Y-axis denotes the proportion of nodes in the failure nodes set (or correctness nodes set) at the respective score level. The results of four attributes are presented in four figures: (a), (b), (c), and (d). The distribution of the final priority scores is shown in Fig. \ref{tab:metric_distribution}e.

We have several observations. In Fig. \ref{tab:metric_distribution}e, we find that as the priority scores increase, the number of failure nodes rapidly increases while the number of correctness nodes decreases. It indicates that GraphRank can effectively distinguish failure and correctness among the unlabeled nodes based on the priority scores.
In Fig. \ref{tab:metric_distribution}a, we observe that the number of correctness nodes rapidly decreases as the metric rises, akin to Fig. \ref{tab:metric_distribution}e. However, the distribution of failure nodes does not align with the result in Fig. \ref{tab:metric_distribution}e. Instead, it exhibits a slow increase followed by a decrease. Meanwhile, the distribution of probabilistic output metric depicted in Fig. \ref{tab:metric_distribution}b is similar to the distribution in Fig. \ref{tab:metric_distribution}a. From the results shown in Fig. \ref{tab:metric_distribution}ab, we infer that the deterministic output and probabilistic output attributes, which are model-aware, have a challenge in identifying the failure nodes but can recognize correctness nodes effectively.

On the contrary, we observe that the number of failure nodes increases rapidly in Fig. \ref{tab:metric_distribution}c, which indicates that the graph node attributes, which are model-agnostic, can identify failure nodes effectively. However, there are still some correctness nodes with high priority scores, which suggests that the graph node attributes may identify these correctness nodes as failures.
In Fig. \ref{tab:metric_distribution}d, although the variations in correctness nodes are insignificant, the number of failure nodes rapidly increases as the degree decreases. This means that the node degree attributes can also detect failure nodes. From the results shown in Fig. \ref{tab:metric_distribution}cd, we conclude that the model-agnostic attributes (e.g., graph node and node degree attributes) can effectively detect failure nodes. This ability can compensate for the drawback of model-aware attributes in identifying failure nodes. Therefore, combining the model-aware and model-agnostic attributes is necessary to overcome the limitation of relying solely on model-aware attributes. 

\noindent
\textbf{Answer to RQ2:} In GraphRank, the deterministic output and probabilistic output attributes can compensate for the limitations of relying solely on model-aware attributes, facilitating the effective testing of GNNs.

\subsection{RQ3: Ablation study}

\textbf{Results:}
The experiment results of RQ3 are presented in Table \ref{tab:ablation} and Table \ref{tab:mutation_imporve}. The ablation study results in Table \ref{tab:ablation} show that the GraphRank$_{Complete}$, which incorporates all components, achieves the highest performance. Compared to the third variant, the complete GraphRank has average improvements by 5.39\% on ATRC. Furthermore, as additional components are gradually incorporated, the performance of GraphRank demonstrated a corresponding improvement. These results show that each of the components in GraphRank is essential; removing any of them leads to a decrease in effectiveness. 

In Table \ref{tab:mutation_imporve}, we observe that leveraging our framework to handle the mutation attributes can improve performance significantly in most situations. The average improvements on GraphPrior and NodeRank are 5.53\% and 5.45\%, respectively. This experiment demonstrates that our framework can enhance overall performance.

\noindent
\textbf{Answer to RQ3:}
The results indicate that using all components is effective. Furthermore, our framework can improve the performance of other methods that rely solely on model-aware attributes, such as GraphPrior and NodeRank.

\begin{table}[t]
\centering
\caption{Ablation study of core components in GraphRank,  ATRC ($\%$)}
\label{tab:ablation}
\begin{tabular}{cl|cccc} \toprule
\textbf{Dataset}                                                          & \textbf{Variant}                & \textbf{EnGCN} & \textbf{SIGN}  & \textbf{GraphSAGE} & \textbf{ClusterGCN} \\ \midrule
\multirow{4}{*}{\begin{tabular}[c]{@{}c@{}}Ogbn-\\ products\end{tabular}} & GraphRank$_{AW}$                & 69.83          & 69.03          & 62.83              & 64.00               \\
                                                                          & GraphRank$_{AW+AG}$             & 72.91          & 68.99          & 64.30              & 64.92               \\
                                                                          & GraphRank$_{AW+AG+EN}$          & 74.19          & 71.92          & 63.01              & 64.97               \\
                                                                          & \textbf{GraphRank$_{Complete}$} & \textbf{82.64} & \textbf{81.47} & \textbf{72.06}     & \textbf{71.74}      \\ \midrule
\multirow{4}{*}{Reddit}                                                   & GraphRank$_{AW}$                & 62.31          & 60.13          & 52.48              & 54.89               \\
                                                                          & GraphRank$_{AW+AG}$             & 64.29          & 59.69          & 55.63              & 56.05               \\
                                                                          & GraphRank$_{AW+AG+EN}$          & 67.34          & 60.84          & 56.01              & 56.25               \\
                                                                          & \textbf{GraphRank$_{Complete}$} & \textbf{69.76} & \textbf{62.59} & \textbf{57.08}     & \textbf{66.13}      \\ \midrule
\multirow{4}{*}{Flickr}                                                   & GraphRank$_{AW}$                & 61.22          & 59.49          & 60.85              & 60.74               \\
                                                                          & GraphRank$_{AW+AG}$             & 64.56          & 60.65          & 62.15              & 61.20               \\
                                                                          & GraphRank$_{AW+AG+EN}$          & 63.42          & 62.16          & 63.03              & 61.60               \\
                                                                          & \textbf{GraphRank$_{Complete}$} & \textbf{70.90} & \textbf{65.17} & \textbf{64.57}     & \textbf{65.31}   \\ \bottomrule  
\end{tabular}
\end{table}

\begin{table}[]
\centering
\caption{The improvement of GraphRank over the state-of-the-art methods (e.g., GraphPrior and NodeRank) when using their attributes as model-aware attributes, values represent the ATRC (\%)}
\label{tab:mutation_imporve}
\begin{tabular}{cc|cccc}
\toprule
\textbf{Dataset}                       & \textbf{Attributes} & \textbf{EnGCN} & \textbf{SIGN} & \textbf{GraphSAGE} & \textbf{ClusterGCN} \\
\midrule
\multirow{2}{*}{{Ogbn-product}} & GraphPrior          & 16.48          & 8.88          & 2.81               & 5.41                \\
                                       & NodeRank            & 7.92           & 9.03          & 6.37               & 9.33                \\
                                       \midrule
\multirow{2}{*}{{Reddit}}       & GraphPrior          & 3.75           & 0.57          & 10.29              & 2.14               \\
                                       & NodeRank            & 5.81           & 2.49          & 12.06              & 4.06                \\
                                       \midrule
\multirow{2}{*}{{Flickr}}       & GraphPrior          & 5.67           & 1.71          & 7.78               & 0.92                \\
                                       & NodeRank            & 4.03           & 3.57          & -0.81              & 1.48                \\
                                       \bottomrule
\end{tabular}
\end{table}

\subsection{RQ4: Model repair analysis}

\textbf{Results:}
The experiment results of RQ4 are presented in Table \ref{tab:repair}. We observe that the accuracy improvements brought by test input prioritization techniques are higher than those brought by random selection. This means that the improvements are not solely ascribed to the expansion of the training set; the meticulous selection of failure nodes also addresses and rectifies some weaknesses of the model. In addition, we observe that the improvement of GraphRank surpasses other techniques in most cases, providing more evidence about its superiority.

\noindent
\textbf{Answer to RQ4:}
The test nodes selected by the test input prioritization method are advantageous for repairing the model. Moreover, GraphRank, which exhibits the most significant improvement in accuracy, demonstrates its efficient testing capabilities.

\begin{table}[thb]
\centering
\caption{The accuracy improvements on the hold-out validation set of target GNN models when trained on the augmented training set ($\%$)}
\label{tab:repair}
\begin{tabular}{clcccc} \toprule
\textbf{Dataset}               & \textbf{Method} & \textbf{EnGCN} & \textbf{SIGN} & \textbf{GraphSAGE} & \textbf{ClusterGCN} \\ \midrule
\multirow{10}{*}{Ogbn-product} & Random          & 0.10           & 0.07          & 0.14               & 0.11                \\
                               & Entropy         & 0.23           & 0.19          & 0.30               & 0.40                \\
                               & Margin          & 0.24           & 0.19          & 0.32               & 0.33                \\
                               & DeepGini        & 0.25           & 0.24          & 0.29               & 0.35                \\
                               & Dropout         & 0.31           & 0.29          & 0.29               & 0.51                \\
                               & DATIS           & 0.23           & 0.35          & 0.38               & 0.34                \\
                               & NNS             & 0.27           & 0.34          & 0.23               & 0.58                \\
                               & GraphPrior      & 0.22           & 0.31          & 0.23               & 0.52                \\
                               & NodeRank        & 0.24           & 0.27          & 0.29               & 0.65                \\
                               & GraphRank       & \textbf{0.48}  & \textbf{0.39} & \textbf{0.44}      & \textbf{0.73}       \\ \midrule
\multirow{10}{*}{Reddit}       & Random          & 0.19           & -0.04         & 0.05               & 0.05                \\
                               & Entropy         & 0.50           & 0.25          & \textbf{0.29}      & 0.09                \\
                               & Margin          & 0.48           & 0.17          & 0.27               & 0.12                \\
                               & DeepGini        & 0.49           & 0.39          & 0.21               & 0.30                \\
                               & Dropout         & 0.53           & 0.31          & 0.24               & 0.17                \\
                               & DATIS           & 0.53           & 0.29          & 0.23               & 0.31                \\
                               & NNS             & 0.53           & 0.27          & 0.19               & 0.13                \\
                               & GraphPrior      & 0.55           & 0.34          & 0.18               & 0.21                \\
                               & NodeRank        & 0.52           & 0.36          & 0.18               & 0.17                \\
                               & GraphRank       & \textbf{0.56}  & \textbf{0.47} & 0.23               & \textbf{0.42}       \\ \midrule
\multirow{10}{*}{Flickr}       & Random          & 0.23           & 0.17          & 0.27               & -0.07               \\
                               & Entropy         & 0.59           & 0.29          & 0.37               & 0.90                \\
                               & Margin          & 0.56           & 0.49          & 0.53               & 0.73                \\
                               & DeepGini        & 0.65           & 0.31          & 0.40               & 0.40                \\
                               & Dropout         & 0.32           & 0.37          & 0.34               & 0.27                \\
                               & DATIS           & 0.61           & 0.45          & 0.30               & 0.35                \\
                               & NNS             & 0.76           & 0.59          & 0.69               & 0.74                \\
                               & GraphPrior      & 0.41           & 0.57          & 0.53               & 0.79                \\
                               & NodeRank        & 0.70           & 0.64          & 0.46               & 0.80                \\
                               & GraphRank       & \textbf{0.81}  & \textbf{0.71} & \textbf{0.78}      & \textbf{1.10}       \\ \bottomrule
\end{tabular}
\end{table}

\subsection{RQ5: Time overhead analysis of GraphRank}

\begin{figure}[t]
    \includegraphics[width=0.7\linewidth]{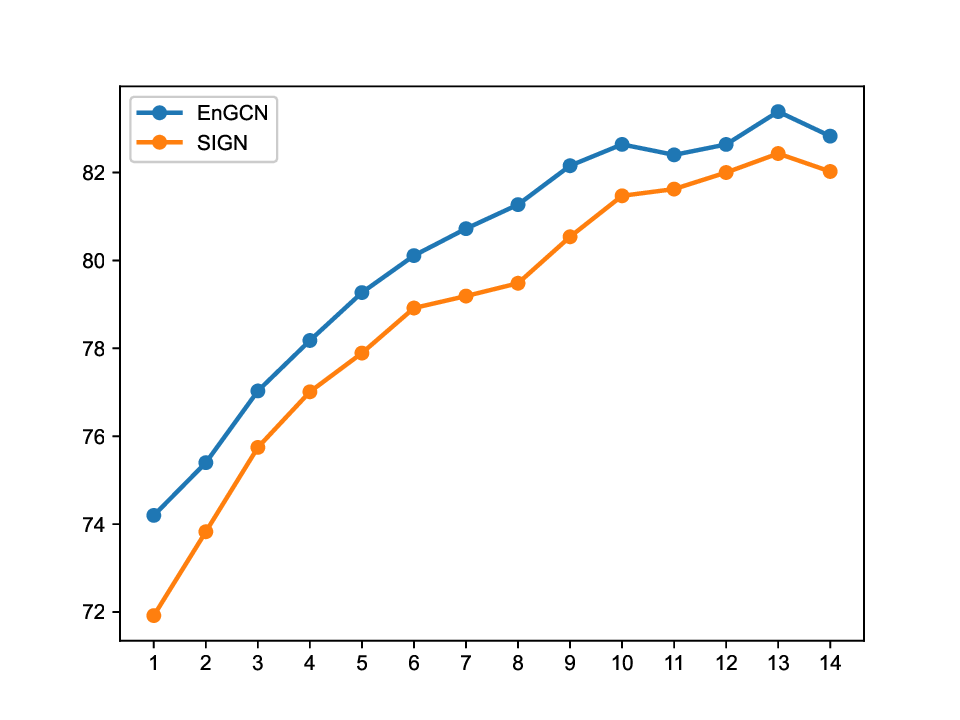}
    \centering
    \caption{The relationship between the number of iteration rounds and GraphRank's performance. X-axis: the number of iteration rounds from 1 to 14. Y-axis: the performance of GraphRank using ATRC ($\%$). We use the Ogbn-products dataset and two GNNs (EnGCN and SIGN) for this experiment}
    \label{fig:ite_tradeoff}
\end{figure}

\textbf{Results:}
The results of RQ6 are reported in Table \ref{tab:time_cost}, and Fig. \ref{fig:ite_tradeoff}
In Table \ref{tab:time_cost}, we present the time overhead of GraphRank and three compared methods. The results indicate that GraphPrior and NodeRank incur significantly higher time overhead compared to GraphRank. This increased overhead is primarily due to the need for these two methods to retrain multiple GNN mutants, which limits their applicability in large-scale graph scenarios. Conversely, while GraphRank requires more time than Dropout due to the training of the MLP and the binary classifier, it achieves a substantial performance improvement of 10.16\% over Dropout. This suggests that GraphRank holds significant potential for practical applications.

We further demonstrate that the costs of training the MLP and the binary classifier in GraphRank are manageable.
First, the MLP, which extracts model-agnostic graph node attributes, requires training only once per graph dataset, making it applicable to various target GNN models of different architectures or hyperparameters. Its simple structure also facilitates easy training. Second, we train the binary classifier iteratively to detect failure nodes. The iterative strategy entails training the binary classifier k times, resulting in a near k-time increase in computational cost. Luckily, we utilize the lightweight machine learning model, i.e., XGBoost~\citep{chen2016xgboost}, as our binary classifier, minimizing the additional time consumption. 

Notably, even a small k can enhance performance, as demonstrated by our analysis of the relationship between k and GraphRank’s effectiveness in Fig. \ref{fig:ite_tradeoff}. With the number of iterations set to 3, the performance of GraphRank on EnGCN and SIGN is 77.03 and 75.74, respectively, surpassing existing test input prioritization techniques.

\noindent
\textbf{Answer to RQ5:}
From the above experiments, we observed that the time overhead of GraphRank is substantially lower than that of the two existing mutation-based methods, indicating that GraphRank has more potential in applications. Moreover, an analysis of the iterative training process reveals that the additional time cost introduced by this component remains well within manageable limits.

\begin{table}[t]
\centering
\caption{The model training and inference time overhead for various test input prioritization methods on the EnGCN model (seconds)}
\label{tab:time_cost}
\begin{tabular}{c c c c | c}
  \toprule
  Dataset & Dropout & GraphPrior & NodeRank & GraphRank  \\
  \midrule
  Ogbn-products & 429.9 & 5487.1  & 5818.9 & 1642.9  \\
  Reddit & 36.0 & 3314.4 & 3368.2 & 136.1 \\
  Flickr & 12.6 & 795.9 & 809.4 & 48.6  \\
  \bottomrule
\end{tabular}

\end{table}

\subsection{RQ6: Performance on small-scale and heterophilic graphs}

\textbf{Results:}
The results of RQ6 are reported in Table \ref{tab:small} and Table \ref{tab:heterophilic}. We observe that our method consistently outperforms the baseline methods on both small-scale and heterophilic datasets. Furthermore, compared to NodeRank, our method achieves an average improvement of 3.15\% and 11.32\% on the small-scale and heterophilic datasets, respectively. These results indicate that our method is effective on both small-scale and heterophilic graph datasets, validating the generalization of GraphRank.

\noindent
\textbf{Answer to RQ6:}
GraphRank can adapt to various types of graph datasets, as it is effective on both large-scale graphs and small-scale or heterophilic graph datasets.

\begin{table}[]
\centering
\caption{Comparing GraphRank with baseline methods using ATRC ($\%$) on small-scale datasets}
\label{tab:small}
\begin{tabular}{cl|ccc} \toprule
\textbf{Dataset}           & \textbf{Method}    & \textbf{GAT}   & \textbf{GCN}   & \textbf{TAGCN} \\ \midrule
\multirow{10}{*}{Cora}     & Random             & 26.51          & 28.48          & 26.82          \\
                           & Entropy            & 54.40          & 66.07          & 45.20          \\
                           & Margin             & 56.13          & 58.29          & 45.20          \\
                           & DeepGini           & 53.74          & 64.80          & 47.43          \\
                           & Dropout            & 53.06          & 63.35          & 45.60          \\
                           & DATIS              & 53.22          & 63.26          & 52.75          \\
                           & NNS                & 53.74          & 64.80          & 47.43          \\
                           & GraphPrior         & 54.29          & 73.94          & 60.48          \\
                           & NodeRank           & 54.10          & 74.76          & 58.53          \\
                           & \textbf{GraphRank} & \textbf{57.03} & \textbf{75.39} & \textbf{61.33} \\ \midrule
\multirow{10}{*}{CiteSeer} & Random             & 8.13           & 20.60          & 11.53          \\
                           & Entropy            & 53.15          & 48.83          & 56.81          \\
                           & Margin             & 52.75          & 54.52          & 58.42          \\
                           & DeepGini           & 53.96          & 49.16          & 58.47          \\
                           & Dropout            & 49.76          & 48.59          & 55.55          \\
                           & DATIS              & 47.35          & 52.19          & 57.78          \\
                           & NNS                & 43.58          & 49.81          & 60.00          \\
                           & GraphPrior         & 54.28          & 48.73          & 62.24          \\
                           & NodeRank           & 50.03          & 53.13          & 64.11          \\
                           & \textbf{GraphRank} & \textbf{54.49} & \textbf{55.56} & \textbf{66.95} \\ \midrule
\multirow{10}{*}{PubMed}   & Random             & 22.65          & 25.09          & 20.37          \\
                           & Entropy            & 50.07          & 78.29          & 47.70          \\
                           & Margin             & 50.49          & 50.65          & 49.93          \\
                           & DeepGini           & 50.37          & 77.99          & 48.08          \\
                           & Dropout            & 49.96          & 77.37          & 47.80          \\
                           & DATIS              & 65.86          & 71.81          & 70.70          \\
                           & NNS                & 70.37          & 77.99          & 78.08          \\
                           & GraphPrior         & 70.70          & 88.45          & 75.49          \\
                           & NodeRank           & 68.10          & 87.35          & 80.30          \\
                           & \textbf{GraphRank} & \textbf{71.66} & \textbf{93.33} & \textbf{82.97} \\ \bottomrule
\end{tabular}
\end{table}

\begin{table}[]
\centering
\caption{Comparing GraphRank with baseline methods using ATRC ($\%$) on heterophilic datasets}
\label{tab:heterophilic}
\begin{tabular}{cl|cc} \toprule
\textbf{Dataset}                & \textbf{Method}    & \textbf{GAT}   & \textbf{Dir-GNN} \\ \midrule
\multirow{9}{*}{Roman-empire}   & Entropy            & 69.34          & 69.41            \\
                                & Margin             & 67.94          & 70.50            \\
                                & DeepGini           & 70.14          & 71.09            \\
                                & Dropout            & 71.01          & 72.21            \\
                                & DATIS              & 63.36          & 67.01            \\
                                & NNS                & 70.14          & 71.09            \\
                                & GraphPrior         & 74.17          & 69.89            \\
                                & NodeRank           & 69.12          & 70.54            \\
                                & \textbf{GraphRank} & \textbf{90.60} & \textbf{87.08}   \\ \midrule
\multirow{9}{*}{Amazon-ratings} & Entropy            & 65.41          & 67.06            \\
                                & Margin             & 64.47          & 65.15            \\
                                & DeepGini           & 66.00          & 67.26            \\
                                & Dropout            & 67.84          & 67.55            \\
                                & DATIS              & 64.50          & 64.73            \\
                                & NNS                & 66.00          & 67.26            \\
                                & GraphPrior         & 68.05          & 64.91            \\
                                & NodeRank           & 65.06          & 66.13            \\
                                & \textbf{GraphRank} & \textbf{69.36} & \textbf{69.08}  \\ \bottomrule
\end{tabular}
\end{table}

\subsection{RQ7: Influence of the binary classifier}

\textbf{Results:}
The results of RQ7 are presented in Table \ref{tab:classifier}. We observe that the performance differences among XGBoost, LightGBM, and Random Forest are marginal, which may be attributed to that the three models are all based on decision trees. These results suggest that other decision tree-based models may also serve as suitable classifiers. In contrast, a notable decline in testing performance is observed when using the MLP and ClusterGCN classifiers, which means that these models may be less suitable for testing. We analyze that the reason could be that the architecture of the MLP is simple, limiting its ability to thoroughly capture priority information. As for the ClusterGCN classifier, since the priority attributes already aggregate high-order neighbor information, further aggregation via ClusterGCN may lead to the over-smoothing~\citep{wang2020direct} problem, ultimately degrading performance.

\noindent
\textbf{Answer to RQ7:}
The experiment results demonstrate that decision tree-based classifiers can achieve better testing performance. However, the MLP and ClusterGCN classifiers experience a significant decrease, indicating that they are less suitable for ranking.

\begin{table}[]
\centering
\caption{Performance of GraphRank without iterative training under different classifiers in terms of ATRC (\%)}
\label{tab:classifier}
\begin{tabular}{cl|cccc} \toprule
\textbf{Dataset}                                                          & \textbf{Classifier} & \textbf{EnGCN} & \textbf{SIGN} & \textbf{GraphSAGE} & \textbf{ClusterGCN} \\ \midrule
\multirow{5}{*}{\begin{tabular}[c]{@{}c@{}}Ogbn-\\ products\end{tabular}} & XGBoost             & 74.19          & 71.92         & 63.01              & 64.97               \\
                                                                          & LightGBM            & 76.56          & 69.88         & 65.08              & 64.24               \\
                                                                          & Random Forest       & 75.17          & 69.87         & 65.46              & 64.89               \\
                                                                          & MLP                 & 64.27          & 68.71         & 57.18              & 57.33               \\
                                                                          & ClusterGCN          & 65.35          & 64.92         & 56.62              & 62.11               \\ \midrule
\multirow{5}{*}{Reddit}                                                   & XGBoost             & 67.34          & 60.84         & 56.01              & 56.25               \\
                                                                          & LightGBM            & 67.88          & 59.36         & 54.09              & 60.18               \\
                                                                          & Random Forest       & 64.49          & 60.35         & 56.38              & 57.51               \\
                                                                          & MLP                 & 49.89          & 53.78         & 46.60              & 45.57               \\
                                                                          & ClusterGCN          & 53.55          & 37.28         &  30.97             & 29.84                \\ \midrule
\multirow{5}{*}{Flickr}                                                   & XGBoost             & 63.42          & 62.16         & 63.03              & 61.60               \\
                                                                          & LightGBM            & 69.52          & 63.76         & 62.84              & 65.62               \\
                                                                          & Random Forest       & 67.75          & 62.09         & 61.15              & 63.76               \\
                                                                          & MLP                 & 57.51          & 51.50         & 52.43              & 54.07               \\
                                                                          & ClusterGCN          & 51.65          & 53.37         & 49.05              & 52.11   \\ \bottomrule           
\end{tabular}
\end{table}

\section{Related work}
\subsection{Prioritization for non-graph models} 
    Most existing test input prioritization techniques are designed for DNNs that process non-graph data (e.g., images). 
    Ma et al.~\citep{ma2018deepgauge} introduce two kinds of testing criteria, namely neuron-level coverage criteria and layer-level coverage criteria. If the coverage of the selected test input set is higher, the set can explore more models' internal states and have a higher chance of revealing failures. 
    Kim et al.~\citep{kim2019guiding} propose two surprise adequacy criteria: Likelihood-based Surprise Adequacy and Distance-based Surprise Adequacy. An input with a higher surprise adequacy value means that it is more difficult to classify by the target model, so these criteria could also be used to estimate priority scores.
    Byun et al.~\citep{byun2019input} propose three prioritization measure methods (confidence, uncertainty, and surprise) to estimate the likelihood of a specific input being a failure.
    Feng et al.~\citep{feng2020deepgini} leverage the Gini metric to measure the dispersion of the distribution. An input has a higher probability of failure if it exhibits a higher degree of dispersion. 
    Li et al.~\citep{li2021testrank} propose TestRank to test DNNs. This method leverages GNNs to extract contextual features, which is beneficial for prioritizing unlabeled test inputs.
    Li et al.~\citep{li2024distance} propose the DATIS metric to select the test inputs. This metric is based on the distances between test inputs and their nearest training samples. 
    Bao et al.~\citep{bao2023defense} propose the NNS metric, which first calculates the $k-$nearest neighbors based on feature representations. Then, it leverages the classification information of both the test sample and its neighbors to compute the priority scores.
    The mutation-based methods~\citep{wang2021prioritizing, dang2024test} defined a series of mutation rules from the model and input mutation to generate different model mutants. Then, a ranking model is employed to predict priority scores based on the difference between the outputs of the target model and the mutants.

    \subsection{Prioritization for graph models} 
Several test input prioritization methods have been developed specifically for Graph Neural Networks (GNNs), including GraphPrior~\citep{dang2023graphprior} and NodeRank~\citep{li2024test}. Both approaches utilize mutation analysis to address this challenge, operating under a core assumption: if a greater number of predictions from mutants differ from the original prediction, the input is more likely to be misclassified and thus should be prioritized for testing.
GraphPrior generates multiple model mutants by adjusting model or training hyperparameters and analyzes their outputs to prioritize test inputs. In contrast, NodeRank employs a similar strategy but enhances it by generating both model mutants and input mutants, thereby increasing the total number of mutants to achieve better performance.
    
    Aside from the above methods, we notice another line of work can be used for test input prioritization -- the ones that focus on estimating the uncertainty of GNNs~\citep{hasanzadeh2020bayesian}. 
    If the target model is uncertain about an input, this input is likely to fail. 
    So the values of uncertainty can be directly used to prioritize graph nodes.
    Dropout~\citep{srivastava2014dropout} is a representative method for uncertainty estimation. 
    It generates multiple outputs for one given input by randomly setting some values to zero in node embeddings matrix $\textbf{H}^l$~\citep{shu2022understanding}.
    Finally, the average of multiple outputs is calculated, and the uncertainty of GNNs is estimated as the information entropy of this average value. 
    In contrast, Graph DropConnect (GDC)~\citep{hasanzadeh2020bayesian} combines three different sampling strategies, namely Dropout, Dropedge~\citep{rong2020dropedge}, and Node Sampling~\citep{chen2018fastgcn}, for uncertainty estimation. 

    \vspace{3pt}
    \noindent
    \textbf{Summary of prior works:} 
Despite the effectiveness of previous approaches, existing methods designed for non-graph models are unable to fully exploit the valuable information embedded in graph structures. Moreover, existing GNN testing methods, i.e., GraphPrior and NodeRank, only use structure information to design mutation rules and generate structure-based attributes. However, in our framework, we not only extract attributes from structural information but also aggregate all attributes through graph structure. After enhancing the attributes, nodes can obtain classification information about neighbors, which proves advantageous for training subsequent ranking models. 

On the other hand, some methods overlook model-agnostic information, leading to incomplete extraction of attributes. TestRank~\citep{li2021testrank} leverages both intrinsic and contextual information to prioritize test inputs. These two types of information correspond to our model-aware and model-agnostic paradigms, respectively. However, the problem arises when it is applied to GNN testing. Since TestRank relies on the GNN model to extract contextual attributes, those attributes become inherently model-aware in GNN testing. This leads to a conflict and makes the method unsuitable for GNN testing.

\section{Discussion}
\subsection{Generalization of GraphRank}
We propose a test input prioritization framework for node classification tasks in GNNs. The evaluation on four large-scale datasets, three small-scale datasets, and two heterophilic graphs demonstrates that our method has the potential for generalization and can be applied to these types of graph datasets.

Additionally, we discuss the potential applicability of GraphRank for other tasks of GNNs. Although our framework is designed for node classification, the underlying concepts of exploring both model-aware and model-agnostic attributes can be extended to other tasks. For example, in edge classification tasks, the features of the two endpoints can serve as model-agnostic attributes. In graph classification tasks, global structural features—such as graph density and node cardinality—may be employed as priority attributes. If appropriate attributes can be designed for other tasks, GraphRank has the potential to be a promising testing method.

\subsection{Threats to Validity}
There are three types of threats to validity: the implementation of GraphRank and the baseline methods, the randomness in the training process, and the selection of the graph datasets and GNN models. To mitigate the first threat, we implement all testing methods by the widely used library \textit{Pytorch}. Additional, we utilize the official open-source codes provided by the original authors for the baseline methods. To mitigate the threats posed by randomness, we conduct four repeated experiments, measuring the stability of GraphRank to reduce this threat.
To address the final threat to validity, we select datasets and models from multiple perspectives, including large-scale, small-scale, and heterophilic.

\section{Conclusion}
\label{sec:conclusion}
In this paper, we present GraphRank, a novel test input prioritization framework specifically designed for GNNs. GraphRank integrates model-aware attributes derived from the GNNs' outputs, alongside model-agnostic attributes such as graph node attributes for improved prioritization performance. Also, we enhance these attributes by aggregating them from neighboring nodes with the graph structure information. These attributes are finally combined with a binary classifier which is iteratively trained to improve the prioritization quality. Empirical results show that GraphRank surpasses existing test input prioritization techniques, heralding a promising direction for future research in this domain.

\section{Declarations}
\subsection{Conflict of intersets}
The authors have no relevant financial or non-financial interests to disclose.

\subsection{Data availability statement}
All datasets used in the experiment are open source. The datasets can be accessed via the following links.

Ogbn-products: \url{https://ogb.stanford.edu/docs/nodeprop/#ogbn-products}

Reddit: 
\url{https://pytorch-geometric.readthedocs.io/en/latest/generated/torch_geometric.datasets.Reddit.html#torch_geometric.datasets.Reddit}

Flickr:
\url{https://pytorch-geometric.readthedocs.io/en/latest/generated/torch_geometric.datasets.Flickr.html#torch_geometric.datasets.Flickr}

\subsection{Code availability statement}
Our code is available at \url{https://github.com/lcyang2/GraphRank}

\section{Acknowledgments}
This work is supported by the National Natural Science Foundation of China (Grant No. 62306093) and Shenzhen Science and Technology Program (Grant No. JSGGKQTD20221101115655027).

\bibliography{sn-bibliography}

\end{document}